\documentclass[aps,prd,nofootinbib,amsmath,amssymb,superscriptaddress,twocolumn,10pt]{revtex4}
\usepackage{graphicx}
\usepackage{dcolumn}
\usepackage{bm}
\usepackage{amssymb}
\usepackage{latexsym}
\usepackage{booktabs}
\usepackage{amsmath}
\usepackage{multirow}
\usepackage[colorlinks=true, linkcolor=red, citecolor=blue]{hyperref}

\def\be{\begin{equation}}
\def\ee{\end{equation}}
\def\ba{\begin{eqnarray}}
\def\ea{\end{eqnarray}}

\newcommand{\potential}{{A}}
\newcommand{\shift}{{B}}
\newcommand{\curvature}{{H}_L}
\newcommand{\shear}{{H}_T}

\newcommand{\ck}{c_K}
\newcommand{\tppi}{\ck p_T \Pi_T}

\begin{document}

\title{Exploring the full parameter space for an interacting dark energy model with recent observations including redshift-space distortions: Application of the parametrized post-Friedmann approach}

\author{Yun-He Li}
\affiliation{Department of Physics, College of Sciences, Northeastern University, Shenyang
110004, China}
\author{Jing-Fei Zhang}
\affiliation{Department of Physics, College of Sciences, Northeastern University, Shenyang
110004, China}
\author{Xin Zhang\footnote{Corresponding author}}
\email{zhangxin@mail.neu.edu.cn}
\affiliation{Department of Physics, College of Sciences,
Northeastern University, Shenyang 110004, China}
\affiliation{Center for High Energy Physics, Peking University, Beijing 100080, China}

\begin{abstract}
Dark energy can modify the dynamics of dark matter if there exists a direct interaction between them. Thus a measurement of the structure growth, e.g., redshift-space distortions (RSD), can provide a powerful tool to constrain the interacting dark energy (IDE) models. For the widely studied $Q=3\beta H\rho_{de}$ model, previous works showed that only a very small coupling ($\beta\sim\mathcal{O}(10^{-3})$) can survive in current RSD data.
However, all of these analyses had to assume $w>-1$ and $\beta>0$ due to the existence of the large-scale instability in the IDE scenario. In our recent work [Phys.\ Rev.\ D {\bf 90}, 063005 (2014)], we successfully solved this large-scale instability problem by establishing a parametrized post-Friedmann (PPF) framework for the IDE scenario. So we, for the first time, have the ability to explore the full parameter space of the IDE models. In this work, we reexamine the observational constraints on the $Q=3\beta H\rho_{de}$ model within the PPF framework. By using the Planck data, the baryon acoustic oscillation data, the JLA sample of supernovae, and the Hubble constant measurement, we get $\beta=-0.010^{+0.037}_{-0.033}$ ($1\sigma$). The fit result becomes $\beta=-0.0148^{+0.0100}_{-0.0089}$ ($1\sigma$) once we further incorporate the RSD data in the analysis. The error of $\beta$ is substantially reduced with the help of the RSD data. Compared with the previous results, our results show that a negative $\beta$ is favored by current observations, and a relatively larger interaction rate is permitted by current RSD data.

\end{abstract}

 \maketitle

\section{Introduction}\label{sec:intro}
Dark energy and dark matter are the dominant sources for the evolution of the current Universe \cite{Ade:2013zuv}. Both are currently only indirectly detected via their gravitational effects. There might, however, exist a direct non-gravitational interaction between them that does not violate current observational constraints. Furthermore, such a dark sector interaction can provide an intriguing mechanism to
solve the ``coincidence problem''~\cite{coincidence,Comelli:2003cv,Zhang:2005rg,Cai:2004dk} and also induce new features to structure formation by exerting a nongravitational influence on dark matter \cite{Amendola:2001rc,Bertolami:2007zm,Koyama:2009gd}.

In an interacting dark energy (IDE) scenario, the energy conservation equations of dark energy and cold dark matter satisfy
\begin{eqnarray}
 \label{rhodedot} \rho'_{de} &=& -3\mathcal{H}(1+w)\rho_{de}+ aQ_{de}, \\
 \label{rhocdot} \rho'_{c} &=& -3\mathcal{H}\rho_{c}+ aQ_{c},~~~~~~Q_{de}=-Q_{c}=Q,
\end{eqnarray}
where $Q$ denotes the energy transfer rate, $\rho_{de}$ and $\rho_{c}$ are the energy densities of dark energy and cold dark matter, respectively,
$\mathcal{H}=a'/a$ is the conformal Hubble expansion rate, a prime denotes the derivative with respect to the conformal time $\tau$, $a$ is the scale factor of the Universe, and $w$ is the equation of state parameter of dark energy.
Several forms for $Q$ have been constructed and constrained by observational data \cite{He:2008tn,He:2009mz,He:2009pd,He:2010im,Boehmer:2008av,Guo:2007zk,Xia:2009zzb,Wei:2010cs,Li:2011ga,Li:2013bya}. The common data sets used in these works are the cosmic microwave background (CMB), the baryon acoustic oscillation (BAO), the type Ia supernovae (SNIa), as well as the Hubble constant measurement. These observations constrain the IDE models mainly by the geometric measurement information, leading to a significant degeneracy between the constraint results of interaction and background parameters.

This degeneracy results from the fact that the IDE model cannot be distinguished from the uncoupled dark energy model in the background evolution since the expansion history of the Universe given by an IDE model and an uncoupled dark energy model can mimic each other by adjusting the values of their free parameters. Fortunately, the dynamics of dark matter can be modified by dark energy in an IDE model, so any observation containing the structure formation information might be a powerful tool to break this degeneracy. Redshift-space distortions (RSD) arising from peculiar velocities of galaxies on an observed galaxy map provide a direct measurement of the linear growth rate $f(a)$ of the large-scale structure formation \cite{Peacock:2001gs,Guzzo:2008ac}. Currently, a number of RSD data are available from a variety of galaxy surveys, such as 6dFGS, \cite{RSD6dF}, 2dFGRS \cite{RSD2dF}, WiggleZ \cite{RSDwigglez}, SDSS LRG DR7 \cite{RSDsdss7}, BOSS CMASS DR11 \cite{Beutler:2013yhm}, and VIPERS \cite{RSDvipers}. These RSD measurements have been used to constrain the IDE models \cite{Honorez:2010rr,Yang:2014gza,yang:2014vza,Wang:2014xca,Yang:2014hea}. For the widely studied $Q=3\beta H\rho_{de}$ model, recent CMB+BAO+SNIa data give $\beta=0.209^{+0.0711}_{-0.0403}$ ($1\sigma$), while the fit result becomes $\beta=0.00372^{+0.00077}_{-0.00372}$ ($1\sigma$) once the RSD data are added to the analysis \cite{Yang:2014gza}. This result shows that a large interaction rate for the $Q=3\beta H\rho_{de}$ model is ruled out by the RSD data.

However, the above results may not reflect the actual preference of the data sets, because the full parameter space cannot be explored in these works, due to the well-known large-scale instability existing in the IDE scenario. The cosmological perturbations will blow up on the large scales for the $Q\propto \rho_{de}$ model with the early-time $w<-1$ or $\beta<0$ \cite{Clemson:2011an,He:2008si} and for the $Q\propto \rho_{c}$ model with the early-time $w>-1$ \cite{Valiviita:2008iv}. So to avoid this instability, one has to assume $w>-1$ and $\beta>0$ for the $Q\propto \rho_{de}$ model and $w<-1$ for the $Q\propto \rho_{c}$ model in the observational constraint analyses. In practice, the $Q\propto \rho_{c}$ model with $w<-1$ is not favored by the researchers, since $w<-1$ will lead to another instability of our Universe in a finite future. Thus, the $Q\propto \rho_{de}$ case with $w>-1$ and $\beta>0$ becomes the widely studied IDE model in the literature.

The large-scale instability arises from the way of calculating the dark energy pressure perturbation $\delta p_{de}$. In the standard linear perturbation theory, dark energy is considered as a nonadiabatic fluid. Thus, $\delta p_{de}$ contains two parts, the adiabatic pressure perturbation in terms of the adiabatic sound speed and the intrinsic nonadiabatic pressure perturbation in terms of the rest frame sound speed. If dark energy interacts with dark matter, then the interaction term $Q$ will enter the expression of the nonadiabatic pressure perturbation of dark energy. For some specific values of $w$ and $\beta$, as mentioned above, the nonadiabatic mode grows fast at the early times and soon leads to rapid growth of the curvature perturbation on the large scales \cite{Valiviita:2008iv}.

However, current calculation of $\delta p_{de}$ may not reflect the real nature of dark energy, since it can also bring instability when $w$ crosses the phantom divide $w=-1$ even for the uncoupled dark energy \cite{Vikman:2004dc,Hu:2004kh,Caldwell:2005ai,Zhao:2005vj}. As it is, finding an effective theoretical framework to handle the cosmological perturbations of dark energy may be a good choice before we exactly know how to correctly calculate $\delta p_{de}$. The simplified version of the parametrized post-Friedmann (PPF) approach \cite{Hu:2008zd,Fang:2008sn} is just an effective framework but is constructed for the uncoupled dark energy models. In our recent work \cite{Li:2014eha}, we established a PPF framework for the IDE scenario. The large-scale instability problem in all the IDE models can be successfully solved within such a generalized PPF framework. As an example, we used the observational data to constrain the $Q=3\beta H\rho_{c}$ model without assuming any specific priors on $w$ and $\beta$. The fit result showed that the full parameter space of this model can be explored within the PPF framework (also see Ref.~\cite{Richarte:2014yva} for a similar follow-up analysis).

In this work, we focus on the widely studied $Q=3\beta H\rho_{de}$ model with a constant $w$. We use the PPF approach to handle its cosmological perturbations. As mentioned above, previous observational constraints on this model have to assume $w>-1$ and $\beta>0$ to avoid the large-scale instability. Within the PPF framework established in Ref.~\cite{Li:2014eha}, we, for the first time, have the ability to explore the full parameter space of this model. So it is of great interest to see how the constraint results change when we let the parameter space of this model fully free. We perform a full analysis on the $Q=3\beta H\rho_{de}$ model by using current observations including the Planck data, the seven data points of BAO, the recent released JLA sample of SNIa, the Hubble constant measurement, and the ten data points of RSD as well. We show that current observations actually favor a negative $\beta$ when $w<-1$ and $\beta<0$ are also allowed. Moreover, with the help of the RSD data, $\beta$ can be tightly constrained, but unlike the previously obtained results $\beta\sim\mathcal{O}(10^{-3})$ in Refs.~\cite{Yang:2014gza,yang:2014vza}, a relatively larger absolute value of $\beta$ (about $\mathcal{O}(10^{-2})$) is favored by the RSD data.

Our paper is organized as follows. In Sec.~\ref{sec:pertur}, we give the general perturbation equations in the IDE scenario. The perturbations of dark matter are given by the standard linear perturbation theory, while those of dark energy are calculated by using the PPF approach established in Ref.~\cite{Li:2014eha}. Some details of the PPF approach as a supplement of Ref.~\cite{Li:2014eha} are also presented in this section. In Sec.~\ref{sec:constraint}, we show how we use the observations to constrain the $Q=3\beta H\rho_{de}$ model, and give a detailed discussion on the fit results. Our conclusions are given in Sec.~\ref{sec:conclusions}. In Appendix \ref{app:fzeta}, we introduce how to calibrate the function $f_\zeta(a)$ of the PPF approach in a specific IDE model. 

\section{Perturbation equations in the IDE scenario}\label{sec:pertur}
\subsection{General equations}
A dark sector interaction in a perturbed Universe will influence the scalar perturbation evolutions. So let us start with the scalar perturbation theory in an FRW universe. The scalar metric perturbations can be expressed in general in terms of four functions, $A$, $B$, $H_L$, and $H_T$~\cite{Kodama:1985bj,Bardeen},
\begin{align}
&\delta {g_{00}} = -a^{2} (2 {\potential}Y),\qquad\delta {g_{0i}} = -a^{2} {\shift} Y_i,  \nonumber\\
& \qquad\delta {g_{ij}} = a^{2} (2 {\curvature} Y \gamma_{ij} + 2 {\shear Y_{ij}}),
\label{eqn:metric}
\end{align}
where $\gamma_{ij}$ denotes the spatial metric and $Y$, $Y_i$, and $Y_{ij}$ are the eigenfunctions of the Laplace operator, $\nabla^2Y=-k^2Y$, and its covariant derivatives, $Y_i=(-k)\nabla_iY$ and $Y_{ij}=(k^{-2}\nabla_i\nabla_j+\gamma_{ij}/3)Y$, with $k$ the wave number.
Similarly, the perturbed energy-momentum tensor can also be expressed in terms of another four functions---energy density perturbation $\delta\rho$, velocity perturbation $v$, isotropic pressure perturbation $\delta p$, and anisotropic stress perturbation $\Pi$,
\begin{align}
& \delta{T^0_{\hphantom{0}0}} =  - { \delta\rho}Y,\qquad\delta{T_0^{\hphantom{i}i}} = -(\rho + p){v}Y^i, \nonumber\\
& \qquad \delta {T^i_{\hphantom{i}j}} = {\delta p}Y  \delta^i_{\hphantom{i}j}
	+ p{\Pi Y^i_{\hphantom{i}j}}.
\label{eqn:dstressenergy}
\end{align}

With the existence of the dark sector interaction, the conservation laws become %\cite{Kodama:1985bj}
\begin{equation}
\label{eqn:energyexchange} \nabla_\nu T^{\mu\nu}_I = Q^\mu_I, \quad\quad
 \sum_I Q^\mu_I = 0,
\end{equation}
where $Q^\mu_I$ is the energy-momentum transfer vector of $I$ fluid, which can be split in general as %\cite{Kodama:1985bj}
\begin{equation}
Q_{\mu}^I  = a\big( -Q_I(1+AY) - \delta Q_IY,\,[ f_I+ Q_I (v-B)]Y_i\big),\label{eq:Qenergy}
\end{equation}
where $\delta Q_I$ and $f_I$ denote the energy transfer perturbation and momentum transfer potential of $I$ fluid, respectively. In a perturbed FRW Universe, Eqs.~(\ref{eqn:energyexchange}) and (\ref{eq:Qenergy}) lead to the following two conservation equations for the $I$ fluid \cite{Kodama:1985bj},
\begin{widetext}
\begin{eqnarray}
 &{\delta\rho_I'}
	+  3\mathcal{H}({\delta \rho_I}+ {\delta p_I})+(\rho_I+p_I)(k{v}_I + 3 H_L')=a(\delta Q_I-AQ_I),\label{eqn:conservation1}\\
    & [(\rho_I + p_I)({{v_I}-{B}})]'+4\mathcal{H}(\rho_I + p_I)({{v_I}-{B}}) -k{ \delta p_I }+ {2 \over 3}k\ck p_I {\Pi_I} - k(\rho_I+ p_I) {A}=a[Q_I(v-B)+f_I],\label{eqn:conservation2}
\end{eqnarray}
\end{widetext}
where $c_K = 1-3K/k^2$ with $K$ the spatial curvature.

\subsection{The PPF framework for the IDE scenario}

Now we discuss the perturbation evolutions for cold dark matter and dark energy, in the comoving gauge, $B=v_T$ and $H_T=0$, where $v_T$ denotes the velocity perturbation of total matters except dark energy. To avoid confusion, we use the new symbols, $\zeta\equiv H_L$, $\xi\equiv A$, $\rho\Delta\equiv\delta\rho$, $\Delta p\equiv\delta p$, $V\equiv v$, and $\Delta Q_I\equiv\delta Q_I$, to denote the corresponding quantities of the comoving gauge except for the two gauge independent quantities $\Pi$ and $f_I$. For cold dark matter, $\Delta p_c=\Pi_c=0$, thus the evolutions of the remaining two quantities $\rho_c\Delta_c$ and $V_c$ are totally determined by Eqs.~(\ref{eqn:conservation1}) and (\ref{eqn:conservation2}). Note that $\Delta Q_{I}$ and $f_{I}$ can be got in a specific IDE model. For dark energy, we need an extra condition on $\Delta p_{de}$ besides $\Pi_{de}=0$ and Eqs.~(\ref{eqn:conservation1}) and (\ref{eqn:conservation2}) to complete the dark energy perturbation system. A common practice is to treat dark energy as a nonadiabatic fluid and to calculate $\Delta p_{de}$ in terms of the adiabatic sound speed and the rest frame sound speed (see, e.g., Ref.~\cite{Valiviita:2008iv}). However, this will induce the large-scale instability in the IDE scenario, as mentioned above.

So we handle the perturbations of dark energy by using the generalized PPF framework established in Ref.~\cite{Li:2014eha}. As shown in Ref.~\cite{Li:2014eha}, the key point to avoid the large-scale instability is establishing a direct relationship between $V_{de} - V_T$ and $V_T$ on the large scales instead of directly defining a rest-frame sound speed for dark energy and calculating $\Delta p_{de}$ in terms of it. This relationship can be parametrized by a function $f_\zeta(a)$ as \cite{Hu:2008zd,Fang:2008sn}
\begin{equation}
\lim_{k_H \ll 1}
 {4\pi G a^2\over \mathcal{H}^2} (\rho_{de} + p_{de}) {V_{de} - V_T \over k_H}
= - {1 \over 3} \ck  f_\zeta(a) k_H V_T,\label{eq:DEcondition}
\end{equation}
where $k_H=k/\mathcal{H}$. This condition in combination with the Einstein equations gives the equation of motion for the curvature perturbation $\zeta$ on the large scales,
\begin{align}
\lim_{k_H \ll 1} \zeta'  = \mathcal{H}\xi - {K \over k} V_T +{1 \over 3} \ck  f_\zeta(a) k V_T.
\label{eqn:zetaprimesh}
\end{align}
On the small scales, the evolution of the curvature perturbation is described by the Poisson equation, $\Phi=4\pi G a^2\Delta_T \rho_T/( k^2\ck)$, with $\Phi=\zeta+V_T/k_H$. The evolutions of the curvature perturbation at $k_H\gg1$ and $k_H\ll1$ can be related by introducing a dynamical function $\Gamma$ to the Poisson equation, such that
\begin{equation}
\Phi+\Gamma = {4\pi Ga^2
\over  k^2\ck} \Delta_T \rho_T
\label{eqn:modpoiss}
\end{equation}
on all scales. Then compared with the small-scale Poisson equation, Eq.~(\ref{eqn:modpoiss}) gives $\Gamma\rightarrow0$ at $k_H\gg1$. On the other hand, with the help of the Einstein equations and the conservation equations as well as the derivative of Eq.~(\ref{eqn:modpoiss}), Eq.~(\ref{eqn:zetaprimesh}) gives the equation of motion for $\Gamma$ on the large scales,
\begin{equation}\label{eq:gammadot}
\lim_{k_H \ll 1} \Gamma'  = S -\mathcal{H}\Gamma,
\end{equation}
with
\begin{align}
S&={4\pi Ga^2
\over k^2 } \Big\{[(\rho_{de}+p_{de})-f_{\zeta}(\rho_T+p_T)]kV_T \nonumber\\
&\quad+{3a\over k_Hc_K}[Q_c(V-V_T)+f_c]+\frac{a}{c_K}(\Delta Q_c-\xi Q_c)\Big\},\nonumber
\end{align}
where $\xi$ can be obtained from Eq.~(\ref{eqn:conservation2}),
\begin{equation}
\xi =  -{\Delta p_T - {2\over 3}\tppi+{a\over k}[Q_c(V-V_T)+f_c] \over \rho_T + p_T}.
\label{eqn:xieom}
\end{equation}

With a transition scale parameter $c_\Gamma$, we can take the equation of motion for $\Gamma$ on all scales to be \cite{Hu:2008zd,Fang:2008sn}
\begin{equation}
(1 + c_\Gamma^2 k_H^2) [\Gamma' +\mathcal{H} \Gamma + c_\Gamma^2 k_H^2 \mathcal{H}\Gamma] = S.
\label{eqn:gammaeom}
\end{equation}

Here we note that the prime in this paper is used to denote the derivative with respect to the conformal time $\tau$ (i.e., $'\equiv d/d\tau$), but in Ref.~\cite{Li:2014eha}, it is defined to be the derivative with respect to $\ln a$ (i.e., $'\equiv d/d\ln a$). This explains why $\mathcal{H}$ appears in Eqs.~(\ref{eq:gammadot}) and (\ref{eqn:gammaeom}) (compared to the corresponding equations in Ref.~\cite{Li:2014eha}).

%Note that $'\equiv d/d\tau$ used in this paper differs from $'\equiv d/d\ln a$ defined in Ref.~\cite{Li:2014eha}, so that $\mathcal{H}$ appears in Eqs.~(\ref{eq:gammadot}) and (\ref{eqn:gammaeom}).

From the above equations, we can find that all of the perturbation quantities relevant to the equation of motion for $\Gamma$ are those of matters except dark energy. So we can solve the differential equation (\ref{eqn:gammaeom}) without any knowledge of the dark energy perturbations. Once the evolution of $\Gamma$ is obtained, we can immediately get the energy density and velocity perturbations,
\begin{align}
&\rho_{de}\Delta_{de} =- 3(\rho_{de}+p_{de}) {V_{de}-V_{T}\over k_{H} }-{k^{2}\ck \over 4\pi G a^{2}} \Gamma,\\ \label{eqn:ppffluid}
& V_{de}-V_{T} ={-k \over 4\pi Ga^2 (\rho_{de} + p_{de}) F} \nonumber \\
&\quad\quad\quad\times\left[ S - \Gamma' - \mathcal{H}\Gamma + f_{\zeta}{4\pi Ga^2 (\rho_{T}+p_{T}) \over k}V_{T}
\right],
\end{align}
with $F = 1 +  12 \pi G a^2 (\rho_T + p_T)/( k^2 \ck)$.

\subsection{The IDE model}\label{subsec:modelequations}

In the following, we get the evolution equations for the specific IDE model under study in this work. To achieve this, we need to construct a covariant interaction form whose energy transfer can reduce to $Q=3\beta H\rho_{de}$ in the background evolution. A simple physical choice is assuming that the energy-momentum transfer is parallel to the four-velocity of dark matter, so that the momentum transfer vanishes in the dark matter rest frame. Then, we have
\begin{equation}\label{eq:covQ}
Q^{\mu}_c  = -Q^{\mu}_{de}=-3\beta H\rho_{de} u^{\mu}_c,
\end{equation}
with the dark matter four-velocity, $$u^\mu_c = a^{-1}\big(1-AY,\,v_cY^i \big),\quad u_\mu^c = a\big(-1-AY,\,(v_c -B)Y_i\big).$$
Comparing Eq.~(\ref{eq:covQ}) with Eq.~(\ref{eq:Qenergy}), we get
 \begin{eqnarray}
&\delta Q_{de}= -\delta Q_c=3\beta H\rho_{de} \delta_{de},\nonumber\\
&f_{de}=-f_c=3\beta H\rho_{de} (v_c-v),\nonumber \\
& Q_{de}=-Q_c=3\beta H\rho_{de},\label{eqn:emtransfer}
 \end{eqnarray}
where we define the dimensionless density perturbation $\delta_I=\delta\rho_I/\rho_I$ for the $I$ fluid. Substituting Eq.~(\ref{eqn:emtransfer}) into Eqs.~(\ref{rhodedot}) and (\ref{rhocdot}), we can obtain the background evolutions of dark energy and dark matter,
\begin{eqnarray}
  &\rho_{de}=\rho_{de0}a^{-3(1+w-\beta)},\\
   &\rho_{c}=\rho_{c0}a^{-3}\left[1+{\beta\over\beta-w}{\rho_{de0}\over\rho_{c0}}\left(1-a^{3\beta-3w}\right)\right],\label{rhocdot}
\end{eqnarray}
where the subscript ``0'' denotes the value of the corresponding quantity at $a=1$ or $z=0$.

For the dark sector perturbation evolutions, we obtain them in the synchronous gauge since most public numerical codes are written in this gauge. The synchronous gauge is defined by $A=B=0$, $\eta=-H_T/3-H_L$, and $h=6H_L$. Then Eqs.~(\ref{eqn:conservation1}) and (\ref{eqn:conservation2}) reduce to
  \begin{eqnarray}
&\delta_c'+kv_c +{h'\over2} ={3\beta{\cal H}\rho_{de}\over\rho_c}(\delta_c-\delta_{de}), \label{eq:dmdensity}\\
&v_c'+{\cal H}v_c=0,\label{eq:dmvelocity}
 \end{eqnarray}
for cold dark matter in the synchronous gauge. From Eq.~(\ref{eq:dmvelocity}), we can see that the momentum transfer vanishes for the IDE model. So there is no violation of the weak equivalence in this model.

To get the dark energy perturbations in the synchronous gauge, we need to make a gauge transformation, since the PPF approach is written in the comoving gauge. The gauge transformation from the synchronous gauge to the comoving gauge is given by \cite{Hu:2008zd}
\begin{eqnarray}
&\rho_I\Delta_I  = \delta_I \rho_I - \rho_I' v_T/k, \label{eq:transdelta}\\
&\Delta p_I  = \delta p_I  - p_I' v_T/k, \label{eq:transdp}\\
&V_I-V_T=v_I-v_T, \label{eq:transv}\\
&\zeta = -\eta - {v_{T}/k_{H}}. \label{eq:transzeta}
\end{eqnarray}
By using Eq.~(\ref{eq:transdelta}), we can obtain $\Phi$ of Eq.~(\ref{eqn:modpoiss}) in terms of $\delta_T$ and $\Gamma$. Then combining Eq.~(\ref{eq:transzeta}) and the gauge relation $V_{T} = k_{H}(\Phi-\zeta)$ \cite{Hu:2008zd}, we can get another useful transformation relation,
\begin{equation}
V_{T} =v_{T}+{4 \pi G a^2\over \mathcal{H}kc_K}\left(\delta_T \rho_T - \rho_T' {v_T\over k}\right) +k_H\eta-k_H\Gamma.\label{eq:transvt}
\end{equation}

With the help of Eqs.~(\ref{eq:transdelta})--(\ref{eq:transvt}), we can rewrite all the equations of PPF approach in terms of the corresponding quantities in the synchronous gauge. For the IDE model under study, we have
\begin{align}
&\delta_{de} =- 3(1+w){v_{de}\over k_{H}}+3\beta{v_T\over k_{H}}-{k^{2}\ck \over 4\pi G a^{2}\rho_{de}} \Gamma,\label{eq:deltadesync}\\
& v_{de}-v_{T} ={-k \over 4\pi Ga^2 \rho_{de}(1+w) F} \nonumber \\
&\quad\quad\quad\times\left[ S - \Gamma' - \mathcal{H}\Gamma + f_{\zeta}{4\pi Ga^2 (\rho_{T}+p_{T}) \over k}(v_{T}+\sigma)
\right],  \label{eq:vdesync}
\end{align}
in the synchronous gauge, where $$\sigma={4 \pi G a^2\over \mathcal{H}kc_K}\left[\delta\rho_T +3(\rho_T+p_T+\beta\rho_{de}) {v_T\over k_H}\right] +k_H\eta-k_H\Gamma.$$
The source term $S$ of Eq.~(\ref{eqn:gammaeom}) can be rewritten in the synchronous gauge as
\begin{align}
S&={4\pi Ga^2
\over k} \Big\{[\rho_{de}(1+w)-f_{\zeta}(\rho_T+p_T)](v_T+\sigma) \nonumber\\
&\quad+\frac{3\beta\rho_{de}}{k_Hc_K}\Big[\xi-\delta_{de}-3(w-\beta) {v_T\over k_H}\Big]\Big\},\nonumber
\end{align}
where
\begin{equation}
\xi =  -{k\delta p_T -p_T'v_T- {2\over 3}k\tppi+3\beta\mathcal{H}\rho_{de}v_T \over k(\rho_T + p_T)}.\label{eq:xi}
\end{equation}
Here note that due to our studied interaction model with $Q$ proportional to $\rho_{de}$, the dark energy density perturbation $\delta_{de}$ occurs in the expression of the source term $S$. Under such circumstance, how can we solve the equation of motion (\ref{eqn:gammaeom}) for $\Gamma$ before $\delta_{de}$ is got from Eq.~(\ref{eq:deltadesync})? For this issue, we can utilize an iteration approach. For example, we can set an initial value for $v_{de}$ and get the value of $\delta_{de}$ from Eq.~(\ref{eq:deltadesync}). Then we can obtain $S$ and solve the differential equation (\ref{eqn:gammaeom}). Finally, we can update the value of $v_{de}$ from Eq.~(\ref{eq:vdesync}) and start another iteration. The convergence speed of this iteration method is proven to be very quick from our tests.

\begin{figure}[htbp]
  \includegraphics[width=8cm]{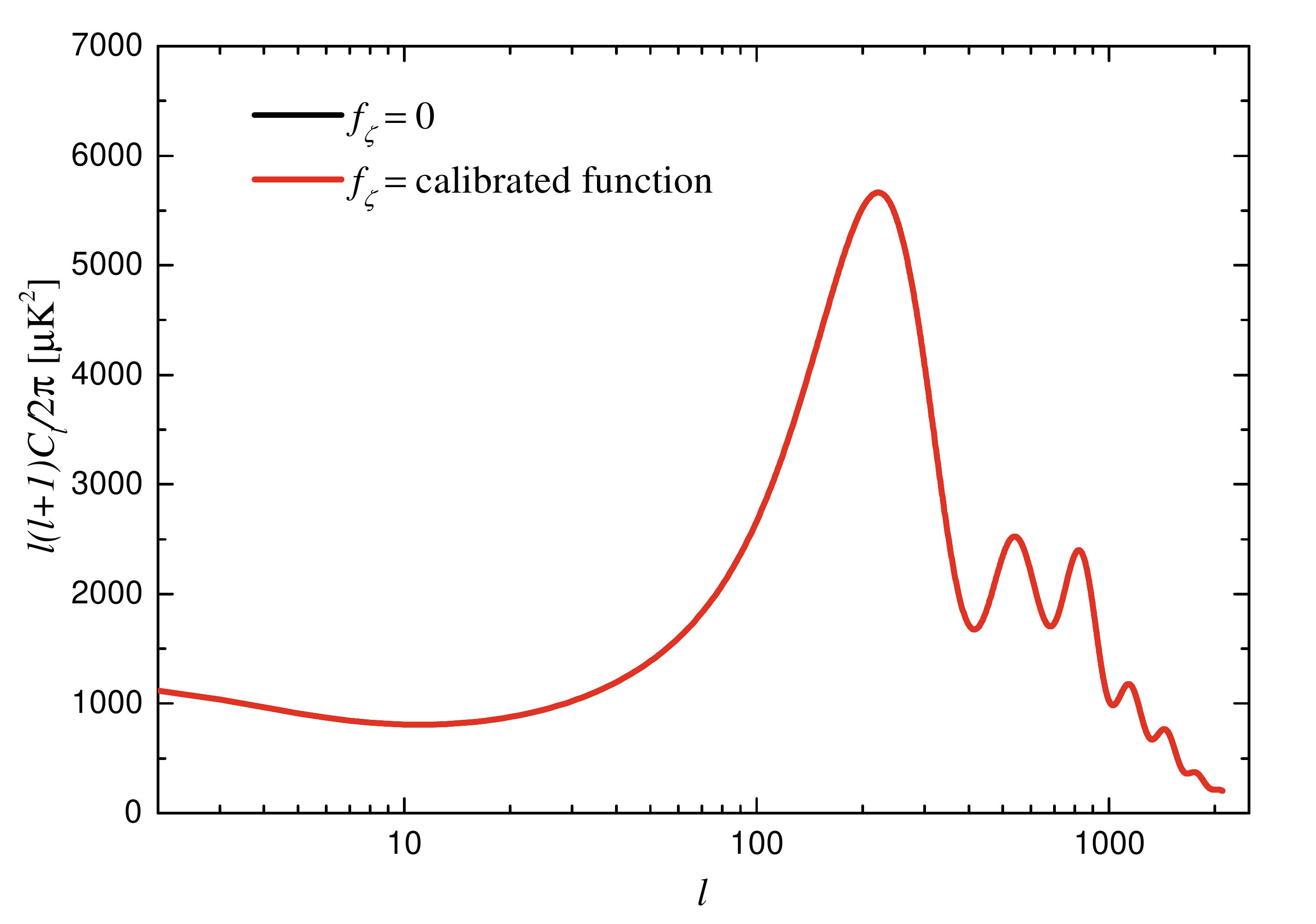}
  \caption{The CMB temperature power spectrum for the $Q^{\mu}=3\beta H\rho_{de}u^{\mu}_c$ model. The red curve denotes the case with $f_\zeta(a)$ taken to be the calibrated function in Eq.~(\ref{eq:fzetafinal}), while the black curve is the case with $f_\zeta(a)=0$. We fix $w=-1.05$, $\beta=-0.01$, and other parameters at the best-fit values from Planck. The overlap of these two curves indicates that the CMB temperature power spectrum does not have the ability to detect the effect of the calibrated $f_\zeta(a)$.}\label{fig:power}
\end{figure}

We also need to determine the parameter $c_\Gamma$ and the function $f_\zeta(a)$.
For the value of $c_\Gamma$, we find that the perturbation evolutions of dark energy are insensitive to its value, so we follow Ref.~\cite{Fang:2008sn} and choose it to be $0.4$. The function $f_\zeta(a)$ can be calibrated in a specific IDE model, but no one gave a concrete way to do this in the previous works. For simplicity, one often takes $f_\zeta(a)=0$ in the literature, because its effect may only be detected gravitationally in the future \cite{Fang:2008sn}. However, we still need, though not urgently, to give a general approach to calibrate $f_\zeta$ for the future high-precision observational data. Besides, we also want to know whether $f_\zeta(a)$ can affect current observations, such as the CMB temperature power spectrum. So we give a detailed process of calibrating $f_\zeta(a)$ in Appendix \ref{app:fzeta}. Using the calibrated $f_\zeta(a)$ given by Eq.~(\ref{eq:fzetafinal}), we plot the CMB temperature power spectrum for the studied IDE model in Fig.~\ref{fig:power}. As a comparison, we also plot the case with $f_\zeta(a)=0$. In this figure, we fix $w=-1.05$, $\beta=-0.01$, and other parameters at the best-fit values from Planck. From Fig.~\ref{fig:power}, we can see that the CMB temperature power spectrum does not have the ability to distinguish these two cases from each other. So it is appropriate to simply set $f_\zeta(a)=0$, currently. Nevertheless, we still utilize the calibrated $f_\zeta(a)$ in our calculations, because we find that taking $f_\zeta(a)$ to be the calibrated function can help to improve the convergence speed of the aforementioned iteration.

\begin{figure*}[htbp]
  \includegraphics[width=8cm]{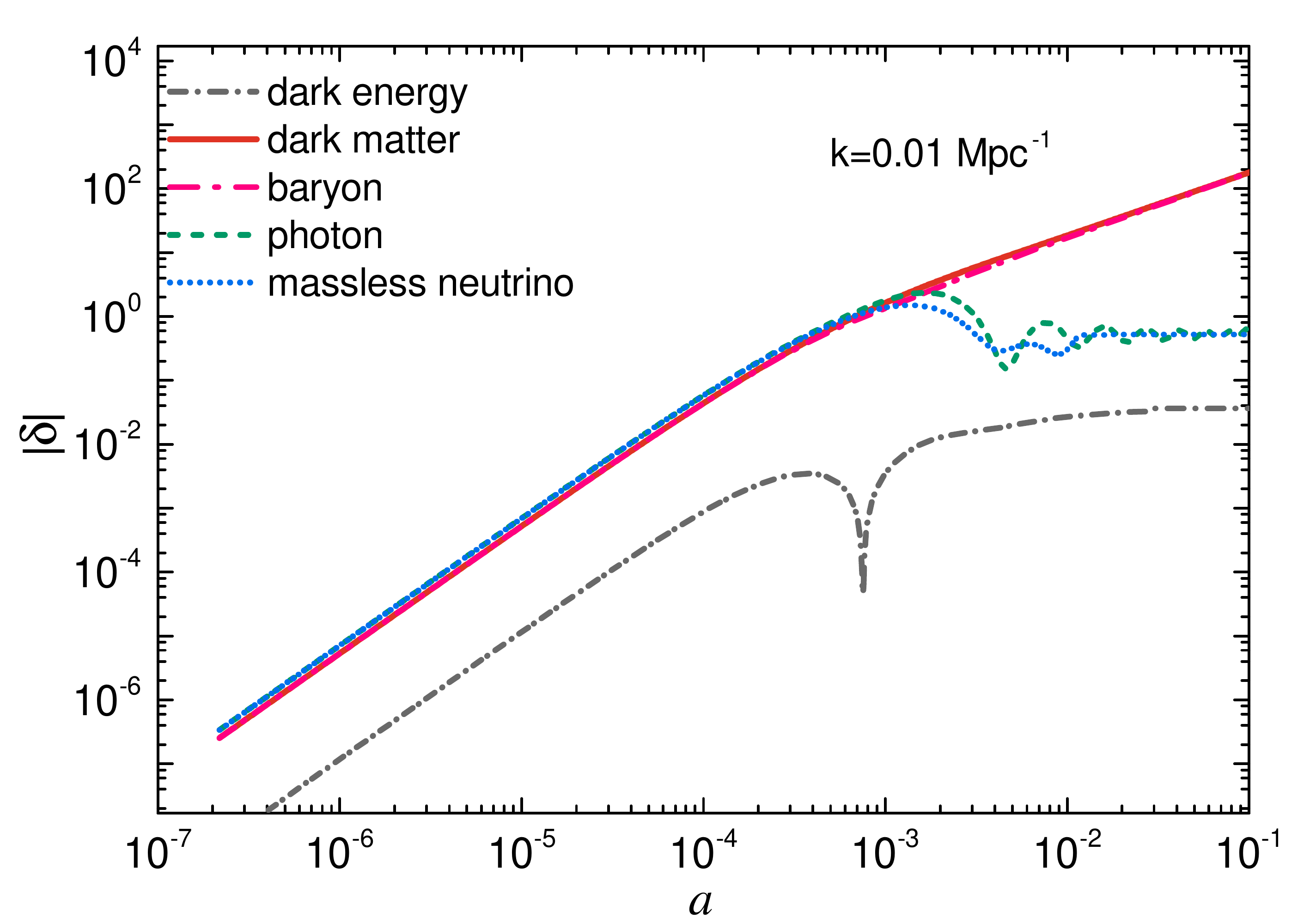}
  \includegraphics[width=8cm]{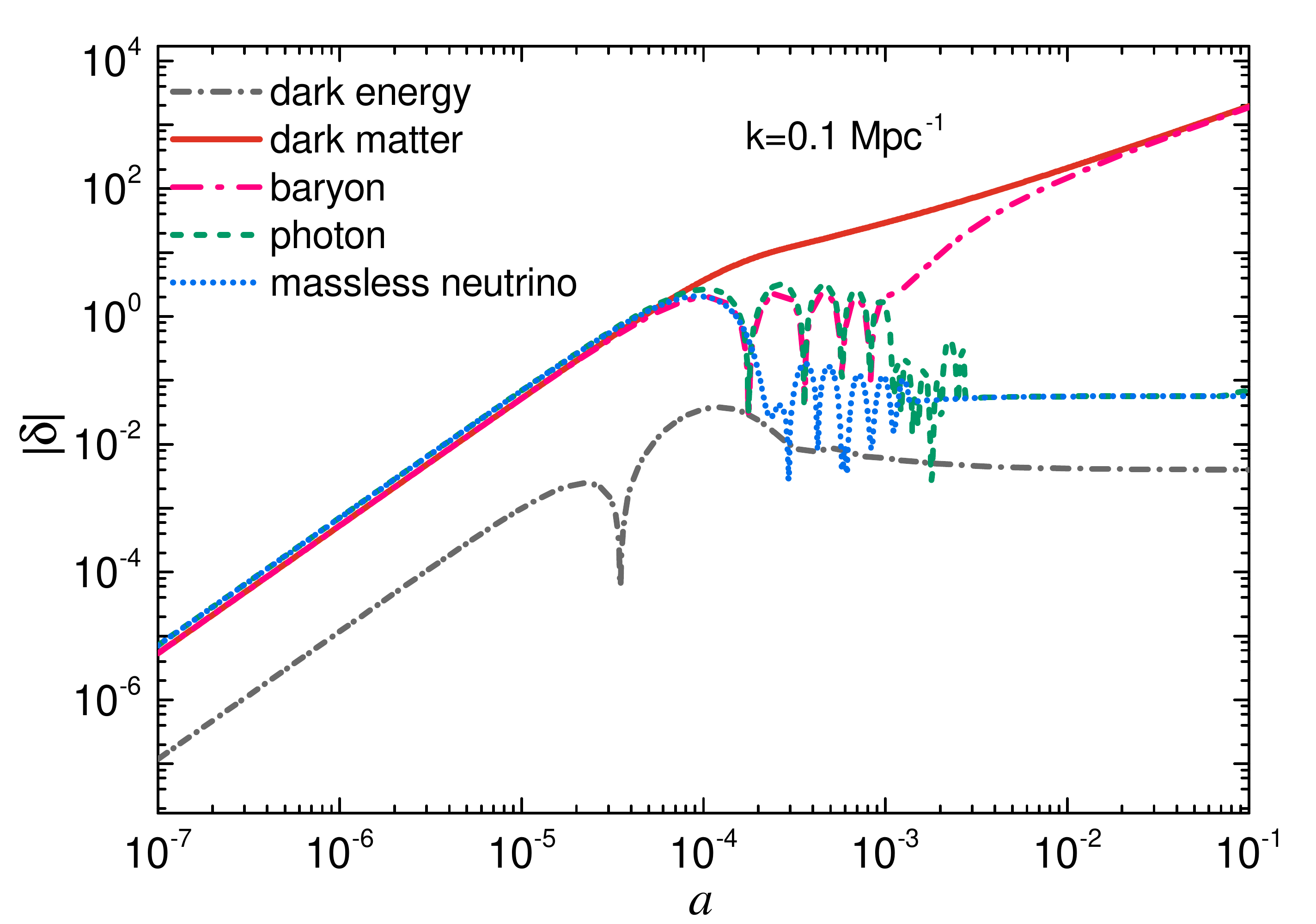}
  \includegraphics[width=8cm]{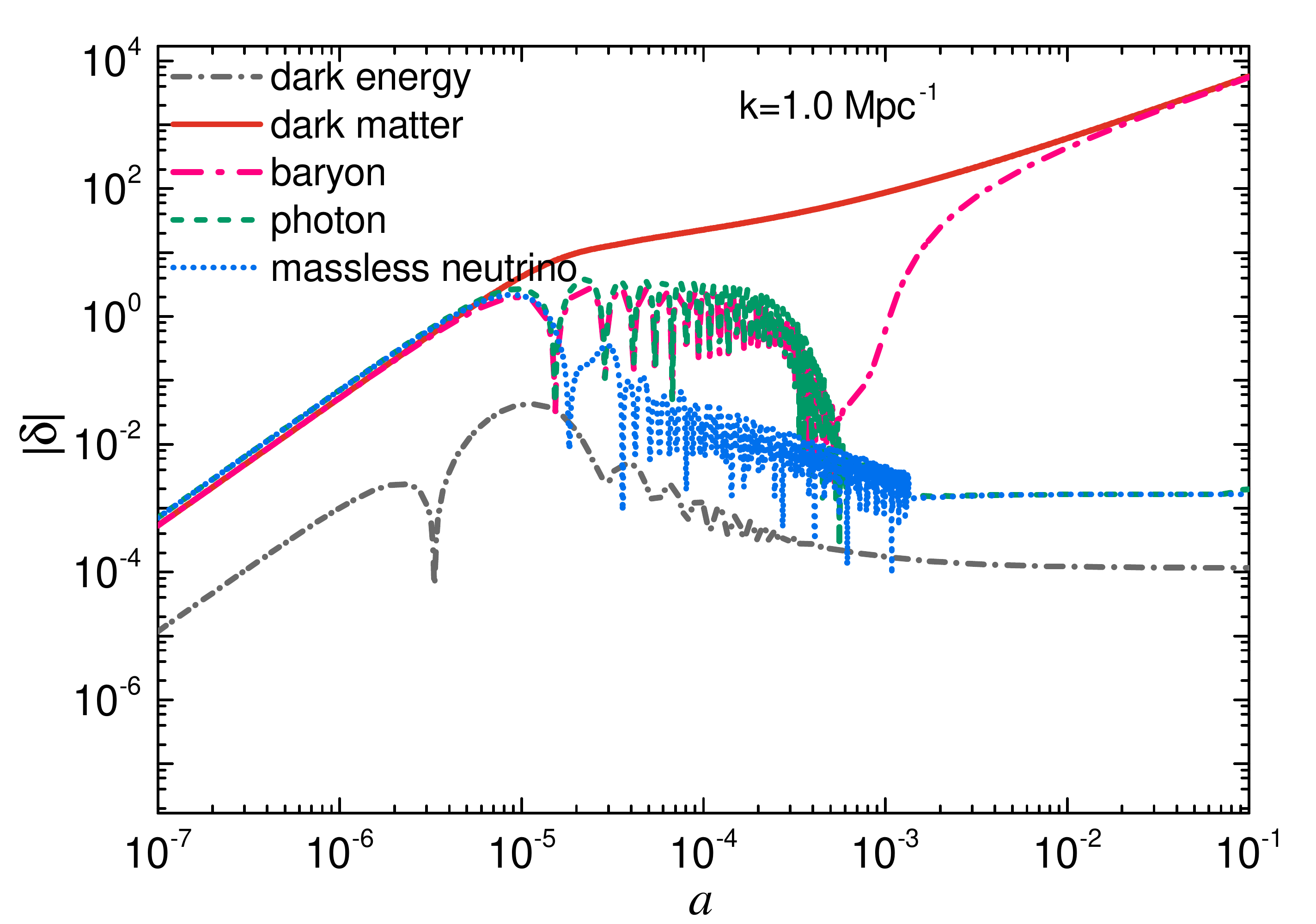}
  \includegraphics[width=8cm]{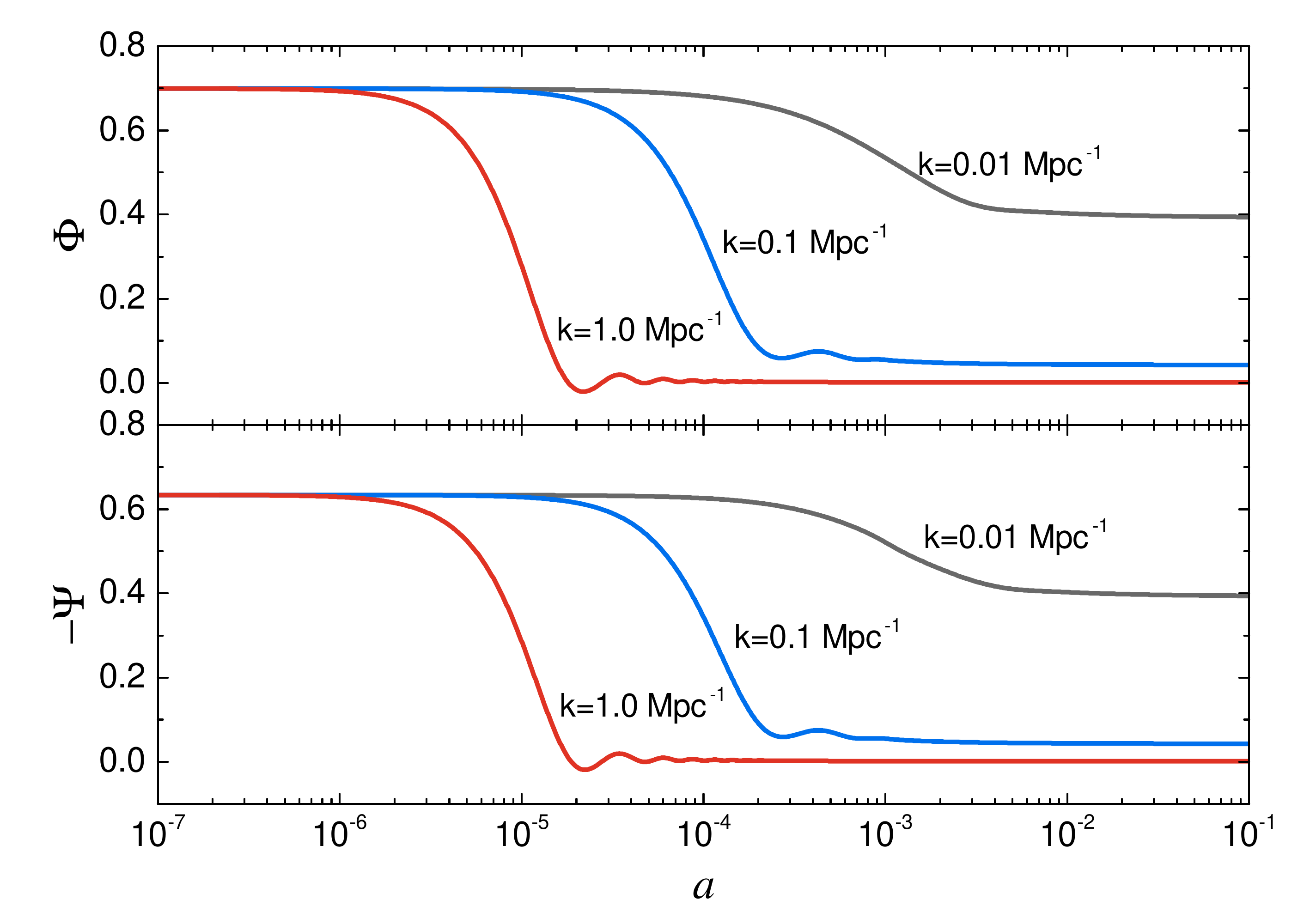}
  \caption{The evolutions of matter and metric perturbations for the $Q^{\mu}=3\beta H\rho_{de}u^{\mu}_c$ model at $k=0.01\,\rm{Mpc^{-1}}$, $k=0.1\,\rm{Mpc^{-1}}$ and $k=1.0\,\rm{Mpc^{-1}}$. Here, the matter perturbations are the corresponding quantities in the synchronous gauge and the metric perturbations $\Phi$ and $\Psi$ are the gauge invariant variables of Kodama and Sasaki \cite{Kodama:1985bj}. We fix $w=-1.05$, $\beta=-0.01$, and other parameters at the best-fit values from Planck. Clearly, all the perturbation evolutions are well behaved.}\label{perturbationevolve}
\end{figure*}

Now, all the perturbation equations can be numerically solved. In Fig.~\ref{perturbationevolve}, we show the matter and metric perturbation evolutions for the IDE model under study at $k=0.01\,\rm{Mpc^{-1}}$, $k=0.1\,\rm{Mpc^{-1}}$ and $k=1.0\,\rm{Mpc^{-1}}$. Here, we also fix $w=-1.05$, $\beta=-0.01$, and other parameters at the best-fit values from Planck. As mentioned above, the $Q^{\mu}=3\beta H\rho_{de}u^{\mu}_c$ model with $w=-1.05$ and $\beta=-0.01$ would be an unstable case if the dark energy perturbations are given by the standard linear perturbation theory. Now we can clearly see from Fig.~\ref{perturbationevolve} that all the perturbation evolutions are stable and normal within the PPF framework.

\section{Data and constraints}\label{sec:constraint}

In this section, we use the latest observational data to constrain the $Q^{\mu}=3\beta H\rho_{de}u^{\mu}_c$ model. We modify the {\tt camb} code \cite{Lewis:1999bs} to include the background and perturbation equations given in Sec.~\ref{subsec:modelequations}. To explore the parameter space, we use the public {\tt CosmoMC} package \cite{Lewis:2002ah}. The free parameter vector is: $\left\{\omega_b,\,\omega_c,\,H_0,\,
\tau,\, w,\, \beta, n_{\rm{s}},\, {\rm{ln}}(10^{10}A_{\rm{s}})\right\}$, where $\omega_b$, $\omega_c$ and $H_0$ are the baryon density, dark matter density and Hubble constant of present
day, respectively, $\tau$ denotes the optical depth to reionization, and ${\rm{ln}}(10^{10}A_{\rm{s}})$ and $n_{\rm{s}}$ are the amplitude and the spectral index of the primordial scalar perturbation power spectrum for the pivot scale $k_0=0.05\,\rm{Mpc}^{-1}$. Here note that we take $H_0$ as a free parameter instead of the commonly used $\theta_{\rm{MC}}$, because $\theta_{\rm{MC}}$ is dependent on a standard noninteracting background evolution. We set a prior [$-0.15$, 0.15] for the coupling constant $\beta$, and keep the priors of other free parameters the same as those used by Planck Collaboration \cite{Ade:2013zuv}. In our calculations, we fix $N_{\rm{eff}}=3.046$ and $\sum m_{\nu}=0$ eV for the three standard neutrino species.

We use the following data sets in our analysis: Planck+WP: the CMB temperature power spectrum data from Planck \cite{Ade:2013zuv} combined with the polarization measurements from 9-year WMAP \cite{wmap9};
BAO: the latest BAO measurements from 6dFGS ($z=0.1$) \cite{6df}, SDSS DR7 ($z=0.35$) \cite{sdss7}, WiggleZ ($z=0.44$, 0.60, and 0.73) \cite{wigglez}, and BOSS DR11 ($z=0.32$ and 0.57) \cite{boss};
JLA: the latest released 740 data points of SNIa from JLA sample \cite{Betoule:2014frx};
$H_0$: the Hubble constant measurement from HST \cite{Riess:2011yx};
RSD: the RSD measurements from 6dFGS ($z=0.067$) \cite{RSD6dF}, 2dFGRS ($z=0.17$) \cite{RSD2dF}, WiggleZ ($z=0.22$, 0.41, 0.60, and 0.78) \cite{RSDwigglez},
SDSS LRG DR7 ($z=0.25$ and 0.37) \cite{RSDsdss7}, BOSS CMASS DR11 ($z=0.57$) \cite{Beutler:2013yhm}, and VIPERS ($z=0.80$) \cite{RSDvipers}.

As mentioned in Sec.~\ref{sec:intro}, RSD actually reflect the coherent motions of galaxies and hence provide information about the formation of large-scale structure. Due to the existence of the peculiar velocities of galaxies the observed overdensity field $\delta_g$ of galaxies in redshift space is enlarged by a factor of $1+f\mu^2/b$ \cite{Kaiser:1987}, where $\mu$ is the cosine of the angle to the line of sight, $b\equiv\delta_g/\delta$ denotes the large-scale bias, and $f(a)\equiv d\ln D(a)/d\ln a$ is the linear growth rate, with the growth factor $D(a)=\delta(a)/\delta(a_{\rm{ini}})$. Thus, through precisely measuring the RSD effect from galaxy redshift surveys, one can obtain information of $f(a)$. However, this measurement of $f(a)$ is bias-dependent. To avoid this issue, Song and Percival \cite{Song:2008qt} suggested using a bias-independent combination, $f(z)\sigma_8(z)$, to extract information from the RSD data, where $\sigma_8(z)$ is the root-mean-square mass fluctuation in spheres with radius $8h^{-1}$ Mpc at redshift $z$. To use the RSD data, we need to do some modifications to the {\tt CosmoMC} package. First, we add an extra subroutine to the {\tt CAMB} code to output the theoretical values of $f(z)\sigma_8(z)$. Here the calculation of $\sigma_8(z)$ inherits from the existing subroutine of the {\tt CAMB} code and $f(a)$ is calculated by $f(a)=d\ln \delta/d\ln a$ with $\delta=(\rho_c\delta_c+\rho_b\delta_b)/(\rho_c+\rho_b)$. Then, we transfer the obtained theoretical values of $f(z)\sigma_8(z)$ to the source files of the {\tt CosmoMC} package and calculate the $\chi^2$ value of the RSD data.

\begin{figure*}[htbp]
\includegraphics[width=17cm]{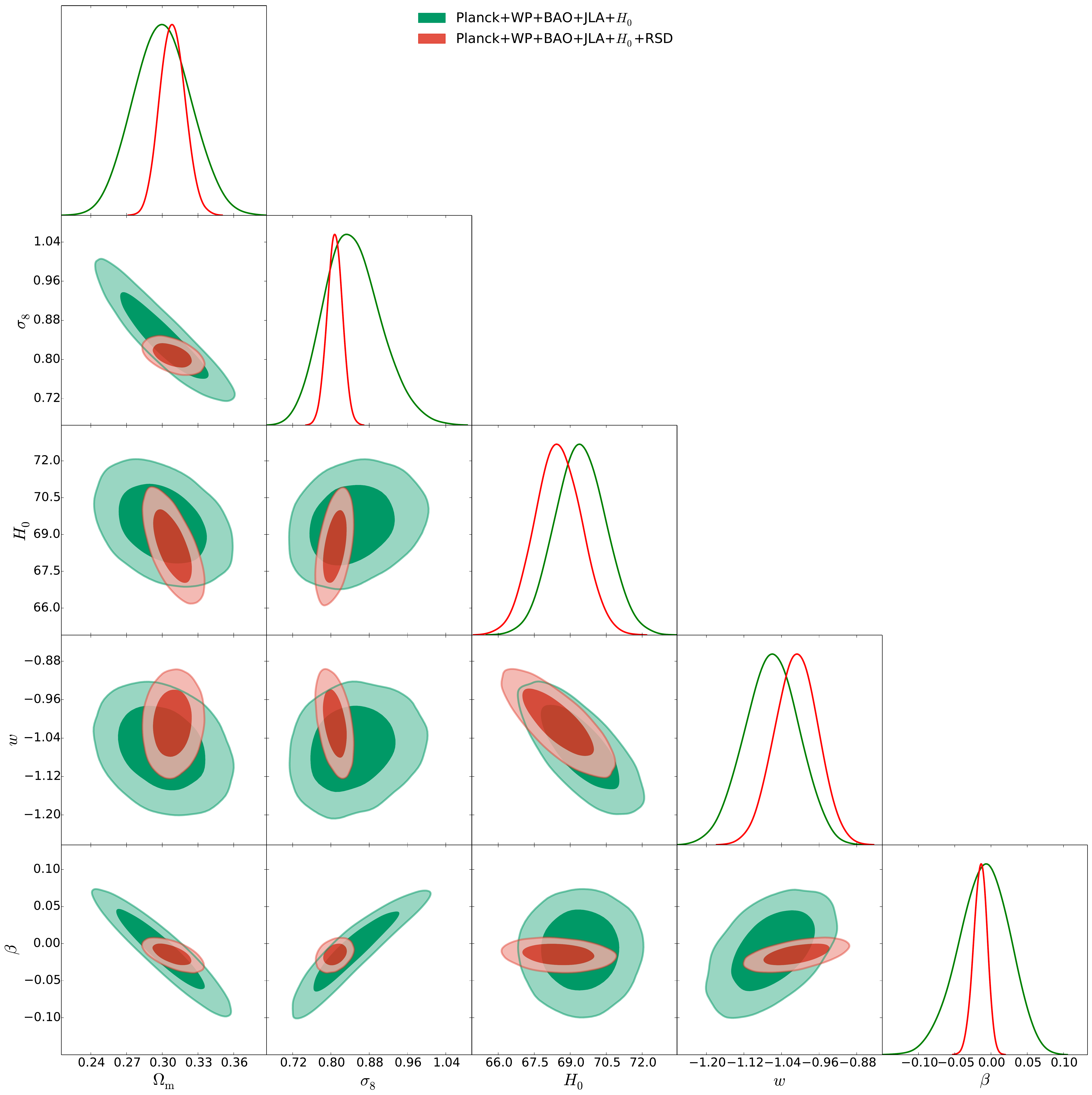}
\caption{The one-dimensional marginalized distributions and two-dimensional marginalized 68.3\% and 95.4\% contours, for the parameters in the $Q^{\mu}=3\beta H\rho_{de}u^{\mu}_c$ model. Within the PPF framework, the full parameter space can be explored. The degeneracy between $\beta$ and $\Omega_m$ is substantially reduced with the help of the RSD data.}\label{contours}
\end{figure*}

First, we constrain the IDE model under study by using the Planck+WP+BAO+JLA+$H_0$ data combination. This data combination can be safely used since the BAO and JLA data are well consistent with the Planck+WP data \cite{Ade:2013zuv,Betoule:2014frx}, and the tension between Planck data and $H_0$ measurement can be greatly relieved in a dynamical dark energy model \cite{Li:2013dha}. The fit results are shown in Table \ref{table1} and Fig.~\ref{contours}. Obviously, the whole parameter space can be explored. By using this data combination, we get $w=-1.061\pm0.056$ and $\beta=-0.010^{+0.037}_{-0.033}$ at the $1\sigma$ level for the $Q^{\mu}=3\beta H\rho_{de}u^{\mu}_c$ model. The parameter space for $w$ and $\beta$ shifts left dramatically compared with the previous results, e.g., $w=-0.940^{+0.0158}_{-0.0599}$ and $\beta=0.209^{+0.0711}_{-0.0403}$ obtained by using a similar data combination but assuming $w>-1$ and $\beta>0$ \cite{Yang:2014gza}. This qualitative change indicates that compulsively assuming $w>-1$ and $\beta>0$ can induce substantial errors on the observational constraint results. So it is of great importance to use the PPF approach to get the correct, whole parameter space for the IDE models.

Although the Planck+WP+BAO+JLA+$H_0$ data combination can give a good constraint result, there still exists a significant degeneracy between the coupling parameter $\beta$ and the background parameter $\Omega_m$ (see the green contours in Fig.~\ref{contours}). This degeneracy, as mentioned in Sec.~\ref{sec:intro}, is hard to break by only using the geometric measurements. So next, we add the extra structure formation information from the RSD data into our analysis. The fit results can also be found in Table \ref{table1} and Fig.~\ref{contours}. By using the Planck+WP+BAO+JLA+$H_0$+RSD data, we get $\beta=-0.0148^{+0.0100}_{-0.0089}$ and $\Omega_m=0.309^{+0.010}_{-0.011}$ at the $1\sigma$ level. Clearly, the errors of $\beta$ and $\Omega_m$ and the degeneracy between them are substantially reduced. Besides, we can also see that current RSD data favor a relatively larger interaction rate for the studied model. This result remarkably differs from that in Ref.~\cite{Yang:2014gza} where a very small positive coupling constant, $\beta=0.00372^{+0.00077}_{-0.00372}$ (1$\sigma$), is obtained by the CMB+BAO+SNIa+RSD data combination. Since $w>-1$ and $\beta>0$ are assumed in Ref.~\cite{Yang:2014gza}, it can be deduced that such a small positive coupling constant just arises from the cut-off effect of the parameter space rather than reflecting the actual preference of the RSD data. So we conclude that our work gives the correct and tightest fit results for the $Q^{\mu}=3\beta H\rho_{de}u^{\mu}_c$ model.

\begin{table}[tbp]
\centering\caption{\label{table1} The mean values and $1\sigma$ errors of all the free parameters and some derived parameters for the $Q^{\mu}=3\beta H\rho_{de}u^{\mu}_c$ model. It can be found that current observations actually favor a negative value of $\beta$. The errors of $\beta$ and $\Omega_m$ are substantially reduced once the RSD data are used.}
\begin{tabular}{lcc}
\hline
\hline
Parameters & Planck+WP+BAO+JLA+$H_0$  & +RSD \\
\hline
$\omega_b$&$0.02209^{+0.00025}_{-0.00026}$&$0.02220^{+0.00025}_{-0.00024}$\\
$\omega_c$&$0.123\pm0.011$&$0.1226^{+0.0039}_{-0.0038}$\\
$H_0$&$69.4\pm1.0$&$68.5^{+1.0}_{-0.9}$\\
$\tau$&$0.089^{+0.012}_{-0.014}$&$0.087^{+0.012}_{-0.013}$\\
$w$&$-1.061\pm0.056$&$-1.009\pm0.045$\\
$\beta$&$-0.010^{+0.037}_{-0.033}$&$-0.0148^{+0.0100}_{-0.0089}$\\
${\rm{ln}}(10^{10}A_s)$&$3.088^{+0.024}_{-0.026}$&$3.079^{+0.023}_{-0.026}$\\
$n_s$&$0.9601\pm0.0057$&$0.9638\pm0.0057$\\
$\Omega_\Lambda$&$0.700\pm0.024$&$0.691^{+0.011}_{-0.010}$\\
$\Omega_m$&$0.300\pm0.024$&$0.309^{+0.010}_{-0.011}$\\
$\sigma_8$&$0.846^{+0.051}_{-0.065}$&$0.808\pm0.016$\\

\hline
$\chi^2_{\rm{min}}$ &10508.090 &10519.498   \\
\hline
\hline
\end{tabular}
\end{table}

Finally, we make a comparison for the $Q=3\beta H\rho_{de}$ model and the $Q=3\beta H\rho_{c}$ model according to their cosmological constraint results. For the $Q=3\beta H\rho_{c}$ model, we got $\beta=-0.0013\pm0.0008$ ($1\sigma$) by using the CMB+BAO+SNIa+$H_0$ data in Ref.~\cite{Li:2014eha}. Then, Ref.~\cite{Richarte:2014yva} further studied this model by adding the RSD data into the analysis. However, the fit result $\xi_c=0.0014\pm0.0008$ ($\xi_c=-\beta$) obtained in Ref.~\cite{Richarte:2014yva} shows that the extra information from the RSD data nearly has no contribution to the fit result of the coupling constant. This phenomenon is somewhat counter-intuitive but still can be understood. Since the coupling form is proportional to $\rho_{c}$ whose value is far greater than that of $\rho_{de}$ at the early times, the amplitude of dark energy perturbation is greatly enlarged by the coupling at the early times (see Fig. 1 of Ref.~\cite{Li:2014eha}), inducing significant effect on the large-scale CMB power spectrum even for a small coupling constant. Thus CMB itself can provide tight constraint on the coupling constant for the $Q=3\beta H\rho_{c}$ model. This feature makes it easy to rule out the $Q=3\beta H\rho_{c}$ model in the future. However, there is no such issue for the $Q=3\beta H\rho_{de}$ model. From this point, we believe that the $Q=3\beta H\rho_{de}$ model should deserve more attention in the future works.

\section{Conclusion}\label{sec:conclusions}
Current astronomical observations provide us with substantial room to study the possible interaction between dark energy and dark matter.
However, such an IDE scenario occasionally suffers from the well-known large-scale instability.
In our previous work \cite{Li:2014eha}, we successfully solved this instability problem by establishing a PPF framework for the IDE scenario for the first time.
However, there are also some issues needing our further discussions. For example, how do we apply the PPF framework to the widely studied $Q\propto\rho_{de}$ model? How do we calibrate $f_\zeta$ in a specific IDE model? More importantly, how will the cosmological constraint results, especially by using the structure formation measurement from RSD, be changed in the widely studied $Q\propto\rho_{de}$ model once $w$ and $\beta$ can be fully let free within the PPF framework?

To answer all of these questions, in this work we focus on the $Q=3\beta H\rho_{de}$ model with the momentum transfer vanishing in the dark matter rest frame. So the covariant interaction form in a perturbed Universe is $Q^{\mu}=3\beta H\rho_{de}u^{\mu}_c$. We handle its cosmological perturbations by using the PPF framework established in Ref.~\cite{Li:2014eha}.
For the problem of how we can solve the equation of motion for $\Gamma$ before the exact value of $\delta_{de}$ is known, we introduce an iteration method. We give a concrete way to calibrate $f_\zeta$. We find that the effect of taking $f_\zeta$ to be the calibrated function cannot be detected by the CMB temperature power spectrum. However, the general calibration approach we provide in this paper may play a crucial role for the future high-precision observations.
Finally, we perform a full analysis on this model with current observations. By using the Planck+WP+BAO+JLA+$H_0$ data combination, we get $w=-1.061\pm0.056$, $\beta=-0.010^{+0.037}_{-0.033}$, and $\Omega_m=0.300\pm0.024$ at the $1\sigma$ level. The fit results become $w=-1.009\pm0.045$, $\beta=-0.0148^{+0.0100}_{-0.0089}$, and $\Omega_m=0.309^{+0.010}_{-0.011}$ once we further incorporate the RSD data into the analysis.

From the above results, we have the following conclusions: (1) Within the PPF framework, the full parameter space of $w$ and $\beta$ can be explored for the $Q^{\mu}=3\beta H\rho_{de}u^{\mu}_c$ model. (2) The fit results show that current observations actually favor a negative coupling constant once $w<-1$ and $\beta<0$ are also allowed by the PPF framework. (3) With the help of the RSD data, the errors of $\beta$ and $\Omega_m$ and the degeneracy between them are substantially reduced. (4) Compared with the previous works, our results show that a relatively larger absolute value of $\beta$ can survive in current RSD data. We believe that our work gives the correct and tightest fit results for the $Q^{\mu}=3\beta H\rho_{de}u^{\mu}_c$ model. The shortage of our work is that we have not considered the $Q^{\mu}\propto u^{\mu}_{de}$ case. In this case, the Euler equation for dark matter is modified, and hence, the weak equivalence principle is broken. However, this breakdown of the weak equivalence principle might be detected by the future weak lensing measurement \cite{Koyama:2009gd}. For current observations, it has been found that the observational constraint results in these two cases are similar \cite{Clemson:2011an}. So we leave this analysis for future work.

\begin{acknowledgments}
We acknowledge the use of {\tt CosmoMC}. This work was supported by the National Natural Science Foundation of China (Grant No.~11175042) and the Fundamental Research Funds for the Central Universities (Grant No.~N120505003).
\end{acknowledgments}

\appendix
\section{Calibration of the function $f_\zeta(a)$}\label{app:fzeta}

The first step to construct the PPF approach is using a function $f_\zeta(a)$ to parametrize the large-scale velocity of dark energy in terms of the total velocity of other matters, as shown in Eq.~(\ref{eq:DEcondition}). This parametrization is based on $V_{de}-V_T={\cal O}(k_H^3 \zeta)$ and $V_T={\cal O}(k_H \zeta)$ at $k_H\ll1$ in the comoving gauge \cite{Hu:2004xd}. Thus, from Eq.~(\ref{eq:DEcondition}), $f_\zeta(a)$ can be calibrated by finding out the exact form of $C(a)\equiv(V_{de}-V_T)/(k_H^2V_T)$ at $k_H\ll1$ in the comoving gauge. Then Eq.~(\ref{eq:DEcondition}) gives
\begin{equation}
f_\zeta(a)=-{12\pi Ga^2 \over c_K\mathcal{H}^2} (\rho_{de} + p_{de})C(a).\label{eq:fzeta}
\end{equation}
In what follows, we show how to get $C(a)$ for the $Q^{\mu}=3\beta H\rho_{de}u^{\mu}_c$ model in detail.

The function $C(a)$ can only be obtained by solving all the standard linear perturbation equations in the comoving gauge where dark energy is treated as a nonadiabatic fluid with its pressure perturbation,
 \begin{equation}
\Delta p_{de} = c_{s}^2\rho_{de}\Delta_{de} +\rho_{de}'(c_{s}^2-c_{a}^2){V_{de}-V_T \over k},\label{eq:deltap}
 \end{equation}
where $c_a$ and $c_s$ are the adiabatic sound speed and rest-frame sound speed of dark energy, respectively. In the following calculations, we take $c_a^2=p_{de}'/\rho_{de}'=w$ and $c_s^2=1$.
Substituting Eq.~(\ref{eq:deltap}) and $\Pi_{de}=0$ into Eqs.~(\ref{eqn:conservation1}) and (\ref{eqn:conservation2}), we get the following two conservation equations for dark energy in the comoving gauge,
\begin{widetext}
  \begin{eqnarray}
&\Delta_{de}'+3{\cal H}(1-w)\Delta_{de}+(1+w)kV_{de}+9{\cal H}^2(1-w^2){V_{de}-V_T \over k} +3(1+w)\zeta'=3\beta{\cal H}\left[3{\cal H}(1-w){V_{de}-V_T \over k}-\xi\right],\label{eq:Deltade}\\
&(V_{de} -V_T)'-2{\cal H}(V_{de} -V_T)-{k\over 1+w}\Delta_{de}-k\xi= {3\beta{\cal H}\over1+w}(V_c+V_T-2V_{de}).
\end{eqnarray}
 \end{widetext}
Similarly, substituting $p_c=\Delta p_c=\Pi_c=0$ into Eqs.~(\ref{eqn:conservation1}) and (\ref{eqn:conservation2}), we obtain
\begin{eqnarray}
&\Delta_c'+kV_c +3\zeta' ={3\beta{\cal H}\rho_{de}\over\rho_c}(\Delta_c-\Delta_{de}+\xi), \\
&(V_c -V_T)'+{\cal H}(V_c -V_T)-k\xi= 0.\label{eq:VcvT}
\end{eqnarray}

The linear perturbation equations of all the components, in principle, are hard to solve analytically. However, since we only focus on the perturbation evolution on the large scales where the period we care about is radiation-dominated one, the perturbation equations can be further simplified and solved analytically. In the early radiation dominated epoch, $\mathcal{H}=\tau^{-1}$, $k_H=k\tau$, $V_b=V_\gamma$ (tight coupling), and $\Pi_I=0$ for $I\neq\nu$, the solutions to the perturbation equations can be found by solving the following first-order differential matrix equation \cite{Doran:2003xq},
\begin{equation}
\label{eq.dUdlnx_gen} \frac{d \bm{U}}{d \ln x} = \mathbf{A}(x) \bm{U}(x),
\end{equation}
where $x=k\tau$, $\mathbf{A}(x)$ is the coefficient matrix and $\bm{U}(x)$ is the matter perturbation vector containing $\Pi_\nu$, $\Delta_I$ for $I=de$, $c$, $\gamma$, $b$, and $\nu$, and $V_I$ for $I=de$, $c$, $\gamma$, and $\nu$. Here the subscripts $\gamma$, $b$, and $\nu$ represent photons, baryons and neutrinos, respectively. As a matter of convenience, we use the following rescaled variables: $\tilde{\Pi}_\nu\equiv\Pi_\nu/x^2$, ${\tilde V}_T\equiv V_T/x$, ${\tilde\Delta}_{I}\equiv\Delta_{I}/x^2$, and ${\tilde V}_{I}\equiv(V_I-V_T)/x^3$ for $I=de$, $c$, $\gamma$, $b$, and $\nu$. Thus, our final matter perturbation vector is
$$\bm{U}^T = \left\{ {\tilde\Delta}_c, \, \tilde{V}_c, \,
{\tilde\Delta}_\gamma, \, \tilde{V}_\gamma, \, {\tilde\Delta}_b, \, {\tilde\Delta}_\nu,\, \tilde{\Pi}_\nu , \, {\tilde\Delta}_{de}, \,
\tilde{V}_{de},  \,\tilde{V}_T \right\},$$
where we solve the differential equation of ${\tilde V}_T$ instead of that of ${\tilde V}_\nu$ so that we can directly get $C(x)={\tilde V}_{de}/{\tilde V}_T$ from the solutions. Note that $(\rho_T+p_T)V_T=\sum_{I=c,\,b,\,\gamma,\,\nu}(\rho_I+p_I)V_I$ and the differential equation for $V_T$ can be found from the second Einstein equation \cite{Hu:2008zd},
\begin{equation}
 V_T'+2\mathcal{H}V_T+k\xi + k\zeta= -{8\pi Ga^2 \over k} {p\Pi}.\label{eq:VTcom}
\end{equation}

Using Eqs.~(\ref{eq:Deltade})--(\ref{eq:VcvT}) and (\ref{eq:VTcom}) and the perturbation equations of photons, baryons and neutrinos given by Ref.~\cite{Doran:2003xq}, we can easily obtain the following evolution equations in terms of the rescaled variables:
\begin{widetext}
\begin{align}
&\frac{d {\tilde\Delta}_{c}}{d\ln x} =-2{\tilde\Delta}_{c} -x^2{\tilde V}_c-{\tilde V}_T +{3\beta\Omega_{de}\over\Omega_c}(x^{-2}\xi+{\tilde\Delta}_{c}-{\tilde\Delta}_{de})-3x^{-2}\frac{d \zeta}{d\ln x},\label{eq.Deltac_newnew}\\     &\frac{d {\tilde V}_c}{d\ln x}  = -4{\tilde V}_c+{\xi\over x^2}, \label{eq.Vc_newnew}\\
&\frac{d {\tilde\Delta}_{\gamma}}{d\ln x} =-2{\tilde\Delta}_{\gamma}
-\frac{4}{3} x^2{\tilde V}_{\gamma}-\frac{4}{3} {\tilde V}_T-4x^{-2}\frac{d \zeta}{d\ln x}, \label{eq.Dg_newnew}\\
&\frac{d{\tilde V}_{\gamma}}{d\ln x} =  \frac{1}{4}
{\tilde\Delta}_{\gamma}  - 3{\tilde V}_\gamma,
\label{eq.Vg_newnew}\\
&\frac{d {\tilde\Delta}_b}{d\ln x} = -2{\tilde\Delta}_b- x^2{\tilde V}_\gamma-{\tilde V}_T-3x^{-2}\frac{d \zeta}{d\ln x},
\label{eq.Db_newnew}\\     &\frac{d {\tilde\Delta}_{\nu} }{d\ln x}=
-2{\tilde\Delta}_{\nu}-\frac{4}{3} x^2{\tilde V}_\nu-\frac{4}{3}{\tilde V}_T-4x^{-2}\frac{d \zeta}{d\ln x},\label{eq.Dn_newnew}\\
&\frac{d {\tilde\Pi}_{\nu}}{d\ln x}
 = \frac{8}{5} x^2{\tilde V}_\nu+\frac{8}{5}{\tilde V}_T-2{\tilde\Pi}_\nu,
 \label{eq.Pnu_newnew} \\
& \frac{d {\tilde\Delta}_{de}}{d\ln x} =-3({5\over3}-w){\tilde\Delta}_{de}-(1+w)x^2{\tilde V}_{de}-(1+w){\tilde V}_T -3(1+w)x^{-2}\frac{d \zeta}{d\ln x}   \nonumber \\
  &\quad\quad\quad~~-9(1-w^2){\tilde V}_{de} +3\beta[-x^{-2}\xi+3(1-w){\tilde V}_{de}],
\label{eq.Dx_newnew} \\
&\frac{d {\tilde V}_{de}}{d\ln x}=-{\tilde V}_{de} +{{\tilde\Delta}_{de}\over 1+w}+x^{-2}\xi+{3\beta\over(1+w)}({\tilde V}_c-2{\tilde V}_{de}),
    \label{eq.Vx_newnew} \\
&\frac{d {\tilde V}_T}{d\ln x}=-3{\tilde V}_T-\Omega_\nu{\tilde \Pi}_\nu-\zeta-\xi,\label{eq:VTEOM}
 \end{align}
where $\Omega_I=8\pi Ga^2\rho_I/(3\mathcal{H}^2)$ denotes the dimensionless energy density of $I$ fluid,
\begin{align}
x^{-2}\xi =  -{\sum_{I=c,\,b,\,\gamma,\,\nu}w_I\Omega_I{\tilde\Delta}_I - {2\over 9}c_K\Omega_\nu{\tilde\Pi}_\nu-3\beta\Omega_{de}{\tilde V}_c \over \sum_{I=c,\,b,\,\gamma,\,\nu}(1+w_I)\Omega_I},
\label{eq:xixeom}
\end{align}
is derived from Eq.~(\ref{eq:xi}), and
\begin{align}
&\zeta=-{\tilde V}_T+ {3\over 2c_K}\left[\sum\Omega_I{\tilde\Delta}_I+3(1+w)\Omega_{de}{\tilde V}_{de}\right],\\
&x^{-2}\frac{d\zeta}{d\ln x} = x^{-2}\xi - {K\over k^2} {\tilde V}_T
- {3\over 2} (1+w)\Omega_{de}{\tilde V}_{de},
\label{eqn:xieom}
\end{align}
can be found from the first and third Einstein equations given in Ref.~\cite{Hu:2008zd}.

In the early radiation dominated epoch, $a=\mathcal{H}_0\sqrt{\Omega_{r0}}\tau$ and $\Omega_I\simeq\rho_I/\rho_r$ with $\rho_r=\rho_\gamma+\rho_\nu$, we have
\begin{align}
 & \Omega_{de} =\frac{\Omega_{de0}}{\Omega_{r0}}\left(\frac{\sqrt{\Omega_{r0}}\mathcal{H}_0}{k}\right)^{1-3(w-\beta)} \, x^{1-3(w-\beta)}, \quad \quad \quad \Omega_{c} =
\left(1+{\beta\over\beta-w}{\Omega_{de0}\over\Omega_{c0}}\right)\frac{\Omega_{c0}}{\sqrt{\Omega_{r0}}}\frac{\mathcal{H}_0}{k} \, x \,, \nonumber\\
&\quad \Omega_{b}=\frac{\Omega_{b0}}{\Omega_{r0}} \, a =\frac{\Omega_{b0}}{\sqrt{\Omega_{r0}}}\frac{\mathcal{H}_0}{k} \, x,\quad \quad \Omega_\nu = {\rho_\nu \over\rho_{r}} = R_\nu, \quad \quad
\Omega_\gamma = 1- \Omega_{b} - \Omega_{c} - \Omega_{de} -\Omega_\nu
\,. \label{eq:Omegas_expl}
\end{align}
Now using Eqs.~(\ref{eq:xixeom})--(\ref{eq:Omegas_expl}), we can obtain the coefficient matrix $\mathbf{A}(x)$ from Eqs.~(\ref{eq.Deltac_newnew})--(\ref{eq:VTEOM}). Furthermore, since $x\ll1$, we can approximate $\mathbf{A}(x)$ by a constant matrix $\mathbf{A}_0$, as long as no divergence occurs when $x\rightarrow0$. For the $w<-1/3$ and small coupling $|\beta|<|w|$ case, there is no divergence term in the matrix $\mathbf{A}_0$ at $x=0$. Thus, we have
\begin{equation}
\mathbf{A}_0=
\left(
\begin{array}{cccccccccc}
 -2 & 0 & \frac{-3}{4} \mathcal{N} & 0 & 0 & \frac{3 R_{\nu }}{4} & -\frac{R_{\nu }}{2} & 0 & 0 & -1 \\
 0 & -4 & \frac{1}{4} \mathcal{N} & 0 & 0 & -\frac{R_{\nu }}{4} & \frac{R_{\nu }}{6} & 0 & 0 & 0 \\
 0 & 0 & -R_{\nu }-1 & 0 & 0 & R_{\nu } & -\frac{2}{3}  R_{\nu } & 0 & 0 & -\frac{4}{3} \\
 0 & 0 & \frac{1}{4} & -3 & 0 & 0 & 0 & 0 & 0 & 0 \\
 0 & 0 & \frac{-3}{4}  \mathcal{N} & 0 & -2 & \frac{3 R_{\nu }}{4} & -\frac{R_{\nu }}{2} & 0 & 0 & -1 \\
 0 & 0 & -\mathcal{N} & 0 & 0 & \mathcal{N}-1 & -\frac{2}{3} R_{\nu } & 0 & 0 & -\frac{4}{3} \\
 0 & 0 & 0 & 0 & 0 & 0 & -2 & 0 & 0 & \frac{8}{5} \\
 0 & 0 & \frac{-3}{4} \mathcal{N}\mathcal{B} & 0 & 0 & \frac{3}{4} \mathcal{B} R_{\nu } & -\frac{1}{2} \mathcal{B} R_{\nu } & 3 w-5 & 9 (w-1) \mathcal{M} & -w-1 \\
 0 & \frac{3 \beta }{w+1} & \frac{1}{4} \mathcal{N} & 0 & 0 & -\frac{R_{\nu }}{4} & \frac{R_{\nu }}{6} & \frac{1}{w+1} & -\frac{w+6 \beta +1}{w+1} & 0 \\
 0 & 0 & \frac{3}{2} \mathcal{N} & 0 & 0 & \frac{-3}{2}  R_{\nu } & -R_{\nu } & 0 & 0 & -2 \\
\end{array}
\right),
\end{equation}
where $\mathcal{N}=R_{\nu }-1$, $\mathcal{B}=w+\beta +1$ and $\mathcal{M}=w-\beta +1$.

The eigenvalues of $\mathbf{A}_0$ can be obtained immediately,
\be
\lambda_i = \left\{0,-4,-3,-2,-2,-2,- \frac{5}{2}   - \frac{\sqrt{1 - 32\,R_\nu
/5}}{2},- \frac{5}{2} + \frac{\sqrt{1 -  32\,R_\nu /5}}{2},
\lambda_{d}^{-}, \lambda_{d}^{+} \right\},
\label{eq.eigenvals}
\ee
where
\be \lambda_{d}^{\pm} =
\frac{-6 + 3\,w}{2}-\frac{3\beta}{1+w} \pm \frac{\sqrt{9w^4+30w^3+13w^2-12\beta w-28w+36\beta^2-12\beta-20}}{2(1+w)}. \label{eq:lambdas}
\ee
The approximate solutions to the matrix equation (\ref{eq.dUdlnx_gen}) can be written as $\sum c_ix^{\lambda_i}\bm{U}_0^{(i)}$ with $c_i$ the dimensionless constant and $\bm{U}_0^{(i)}$ the eigenvector corresponding to eigenvalue $\lambda_i$. Obviously, the mode with negative ${\rm Re}(\lambda_i)$ will soon decay or oscillate. For our studied IDE model, if $w<-1$ or $\beta<0$, ${\rm Re}(\lambda_{d}^{\pm})\gg1$, the curvature perturbation will grow rapidly at the early times, leading to the large-scale instability. So we actually calibrate $f_\zeta(a)$ for the stable $w>-1$ and $\beta>0$ case. When $w>-1$ and $\beta>0$, the only largest eigenvalue in Eq.~(\ref{eq.eigenvals}) is zero, and the corresponding eigenvector is
$$\bm{U}_0^T = \left\{ -\mathcal{P},\,\frac{1}{6}\mathcal{P}-\frac{1}{12},\,-\frac{4}{3}\mathcal{P},\,-\frac{1}{9}\mathcal{P},\,-\mathcal{P},-\frac{4}{3}\mathcal{P},\,\frac{4}{5},{\tilde\Delta}_{de}^{(0)},\,
\tilde{V}_{de}^{(0)},\,1
\right\},$$
where $\mathcal{P}=(2 R_{\nu }+5)/5$ and
$$\tilde{V}_{de}^{(0)}=\frac{4 R_{\nu } \left(-3 \beta +12 w^2+9 \beta  w+4 w-8\right)+5 \left(-3 \beta +12 w^2+9 \beta  w+16 w+4\right)}{60 \left(-21 \beta +12 w^2+9 \beta  w-2 w-14\right)}.$$
This mode dominates the cosmological perturbation evolutions on the large scales, so from Eq.~(\ref{eq:fzeta}) we have
\begin{equation}
f_\zeta(a)=-{12\pi Ga^2 \over c_K\mathcal{H}^2} (\rho_{de} + p_{de})\tilde{V}_{de}^{(0)}.\label{eq:fzetafinal}
\end{equation}
Here note that $\tilde{V}_T^{(0)}=1$.
\end{widetext}

\end{document}